%% file: Main.tex
\documentclass[journal]{IEEEtran}
\usepackage{cite}
\usepackage{xcolor}
\usepackage[pdftex]{graphicx}
\graphicspath{{../pdf/}{../jpeg/}}
\DeclareGraphicsExtensions{.pdf,.jpeg,.png}
\usepackage{graphicx}
\usepackage{hyperref}
\hypersetup{
    colorlinks=true,
    linkcolor=blue,
    filecolor=blue,      
    urlcolor=black,
    citecolor =blue,
    pdftitle={Design of perfect anomalous reflector},
    pdfpagemode=FullScreen
    }

\usepackage{subfigure}
\usepackage[cmex10]{amsmath}
\usepackage{array}
\usepackage{textcomp}
\usepackage{gensymb}
\usepackage{multirow}
\usepackage{multicol}
\usepackage{stfloats}
\usepackage[nolist]{acronym}

\begin{document}

\title{Analysis of Scalable Anomalous Reflectors through Ray Tracing and Measurements}

\author{
Le~Hao,~\IEEEmembership{Student 
Member,~IEEE,}
Sravan~K.~R.~Vuyyuru,~\IEEEmembership{Member,~IEEE,}
Sergei~A.~Tretyakov,~\IEEEmembership{Fellow,~IEEE,}\\
Markus~Rupp,~\IEEEmembership{Fellow,~IEEE,}
and Risto~Valkonen,~\IEEEmembership{Member,~IEEE}

\thanks{This work was supported in part by the European Union’s Horizon 2020 MSCA-ITN-METAWIRELESS project, under the Marie Skłodowska-Curie grant agreement No 956256 and the Research Council of Finland, grant 345178. \textit{(Corresponding author: Le~Hao)}} 
\thanks{L. Hao and M. Rupp are with TU Wien, Gusshausstrasse 25, 1040 Vienna, Austria. (e-mail: \{le.hao, markus.rupp\}@tuwien.ac.at).}
\thanks{S.~K.~R. Vuyyuru is with Nokia Bell Labs, Karakaari 7, 02610 Espoo, Finland
and the Department of Electronics and Nanoengineering, School of Electrical Engineering, Aalto University, 02150 Espoo, Finland (e-mail: sravan.vuyyuru@nokia.com; sravan.vuyyuru@aalto.fi).}
\thanks{S.~A. Tretyakov is with the Department of Electronics and Nanoengineering, School of Electrical Engineering, Aalto University, 02150 Espoo, Finland (e-mail: sergei.tretyakov@aalto.fi).}
\thanks{R. Valkonen is with Nokia Bell Labs, Karakaari 7, 02610 Espoo, Finland (e-mail: risto.valkonen@nokia-bell-labs.com).}
}  

\maketitle

\begin{abstract}
In this study, we elaborate on the concept of scalable \ac{AR} to analyze the angular response, frequency response, and spatial scalability of a designed \ac{AR} across a broad range of angles and frequencies. We utilize theoretical models and ray tracing simulations to investigate the communication performance of two different-sized scalable finite \ac{AR}s, one smaller configuration with $48\times48$ array of unit cells and the other constructed by combining four smaller ARs to form a larger array with $96\times96$ unit cells.
To validate the developed theoretical approach, we conducted measurements in an auditorium to evaluate the received power through an AR link at different angles and frequencies. In addition, models of scalable deflectors are implemented in the MATLAB ray tracer to simulate the 
measurement scenario. The results from theoretical calculations and ray tracing simulations achieve good agreement with measurement results. 
\end{abstract}


\IEEEpeerreviewmaketitle

\input{Acronyms.tex}

\section{Introduction}
\label{sec:Intro}

\IEEEPARstart{A}{nomalous} reflectors with dense two-dimensional grids of subwavelength unit cells have gained attention in realizing deflection of incident waves towards desired directions~\cite{asadchy2016perfect, Rubio2017, vuyyuru, Mostafa2023, Rubio2021EuCAP, Raptis2023,wong_prx2018,kwon_ieeejawpl2018, Rubio2021,vuyyuru23array,asadchy_prx2017,vuyyuru2024, Hao2024}. The \ac{EM} properties of each unit cell are controlled to redirect the impinging waves in the arbitrary desired direction. The initial research on \ac{AR}s using conventional methods~\cite{huang2008, yu_science2011} suffers from lower efficiencies for anomalous reflection angles beyond $45\degree$, e.g.~\cite{asadchy2016perfect}. Extensive research on \ac{AR}s established advanced design methods using periodic reflectors, such as highly efficient anomalous reflectors at a steep angle, e.g.~\cite{Rubio2017,vuyyuru} or multichannel reflection~\cite{asadchy_prx2017, vuyyuru2024}.

Even though extensive research on ARs has been carried out in the past years, current researches usually focus on the electromagnetic parameters of ARs, such as the anomalous reflection efficiency and scattering pattern, but significantly lacking in comprehensive analysis of communication aspects in realistic environments~\cite{Jiang23,Jian22,Huang22,TangmmWave22}. More studies are needed to find out how to incorporate these models into network simulators. Contrarily, the majority of research communication aspects of the use of AR are based on simplified or even idealized models that have not been validated for realistic ARs that are designed using electromagnetic-theory methods~\cite{Najafi21,Ozdogan20}. There is also a deficiency in research addressing the frequency selectivity and spatial scalability of ARs. In this paper, we address these gaps by integrating the electromagnetic parameters of an EM-designed and manufactured AR into different communication models and software that are often used in communication system modeling. We theoretically evaluate large-scale fading of AR-assisted communication links and utilize a ray tracer to simulate realistic indoor scenarios involving an anomalous reflector. The theoretical predictions and ray tracing model simulations have been compared and verified by empirical measurements. 

To the best knowledge of the authors, this is the first work that analyzes the angular response and frequency response of a manufactured  3-bit AR in the communication models through ray tracing and measurements in a dense indoor scenario. In addition, we analyze the size scalability of the manufactured ARs by tiling multiple small AR panels into a larger-size AR. Since the implementation of the AR in a ray tracer has been verified by simulation and measurements, this method can be used for any type of AR. As long as the scattering pattern of the AR is available from analytical estimations, simulations, or test measurements, we can use a ray tracer to obtain very accurate results for communication links in complex environments without the need to perform measurements in every particular setting.

The remainder of this article is organized as follows: Section~\ref{sec:ARintro} briefly introduces the designed and manufactured AR, as well as its validated scattering pattern. Section~\ref{sec:MeasSystem} is devoted to the measurement system, including the measurement scenario and setups. In Section~\ref{sec:Results}, we present the comparison of the results between the theoretical model, the ray tracing simulations, and measurements for two ARs of different sizes. Finally, conclusions are formulated in Section~\ref{sec:Conclusion}.


\section{Scalable Anomalous Reflector: Design and Manufacturing Strategy}
\label{sec:ARintro}

Following the methodology proposed in~\cite{vuyyuru2024, Hao2024} for modeling \ac{AR} from \ac{EM} and communication-related aspects, we employed the same design synthesis and manufacturing strategy to develop scalable ARs. In \cite{vuyyuru2024}, we introduce a reconfigurable anomalous reflector enabled by an algebraic array antenna scattering synthesis technique~\cite{vuyyuru} to optimize the scattering characteristics of passively loaded periodic arrays of patch elements. Using the load impedances as the optimization variables for the scattering synthesis problem replaces heavy computations with supercell-level optimization. As proposed in~\cite{vuyyuru2024}, the flat multimode electrically large 16-element supercell engineers the response of the excited fixed normal incident wave into nine anomalous reflection directions, i.e., allowing reconfigurable reflected Floquet harmonics up to fourth order with an infinite periodic array. The Floquet expansion of currents induced on a periodic metasurface with no parasitic scattering and the supercell dimension determined from Floquet-Bloch theory is $d = 4\lambda/|\sin\theta^{r,\text{max}}|$, with the metasurface element periodicity being $d/16$, where the maximum desired deflection direction is $\theta^{r,\text{max}} = 65\degree$. Following~\cite{vuyyuru2024}, we utilize the same synthesized load parameters to reflect the normal incidence wave to the $+65\degree$ anomalous direction at the center frequency of $26$~GHz.

Recalling the manufacturing process outlined in~\cite{vuyyuru2023finite,vuyyuru2024}, we built a passive and fixed-response anomalous reflector prototype deflecting a normal incident wave to a 65-degree anomalous angle. Each square structured unit cell with dimensions of $\Delta x \times \Delta y = d/16 \times d/16 = 0.2758\lambda \times 0.2758\lambda$ are linearly arranged in the $x$-dimension, resulting in the 16-element supercell size being $4.4135\lambda \times 0.2758\lambda$ in $(x\times y)$ dimensions.
Each unit cell is constructed by a copper square patch applied on top of a printed circuit board (PCB) which uses Rogers RO4350B (LoPro) laminate with a $0.55$~mm thickness ($\epsilon_r= 3.66$, $\tan\delta = 0.0037$). Each unit cell within a supercell is loaded with a reactive impedance implemented by shorted or open coplanar lines according to the simulation-based optimization described in \cite{vuyyuru2024}. A 3-bit quantization resolution for the reactive loads is assumed sufficient, based on the analysis in~\cite{vuyyuru2024, Hao2024}.

\begin{figure}[t]
\centering		
\subfigure[\label{Fig:1a}]{\includegraphics[width=0.242\textwidth]{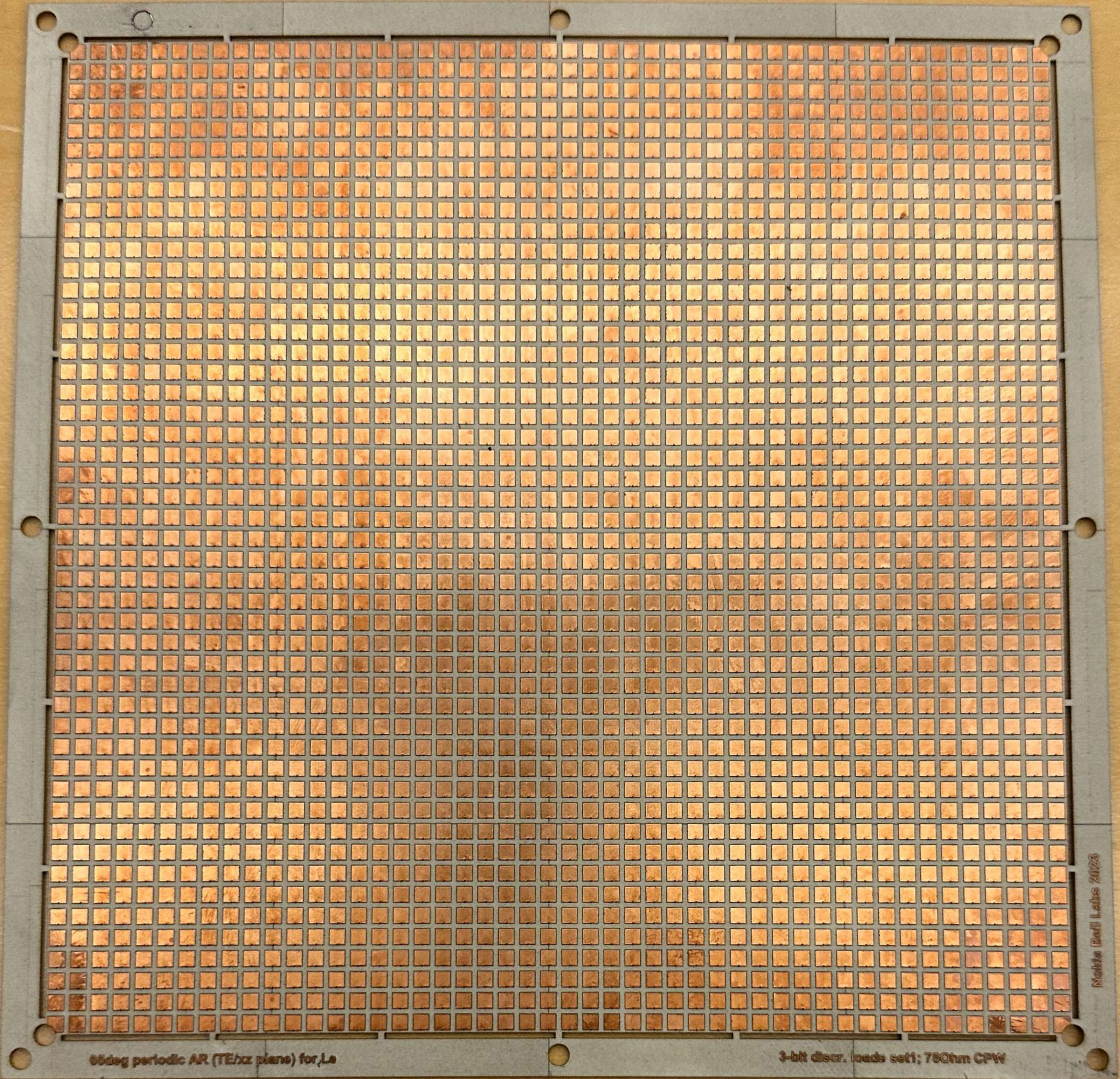}}
\subfigure[\label{Fig:1b}]{\includegraphics[width=0.238\textwidth]{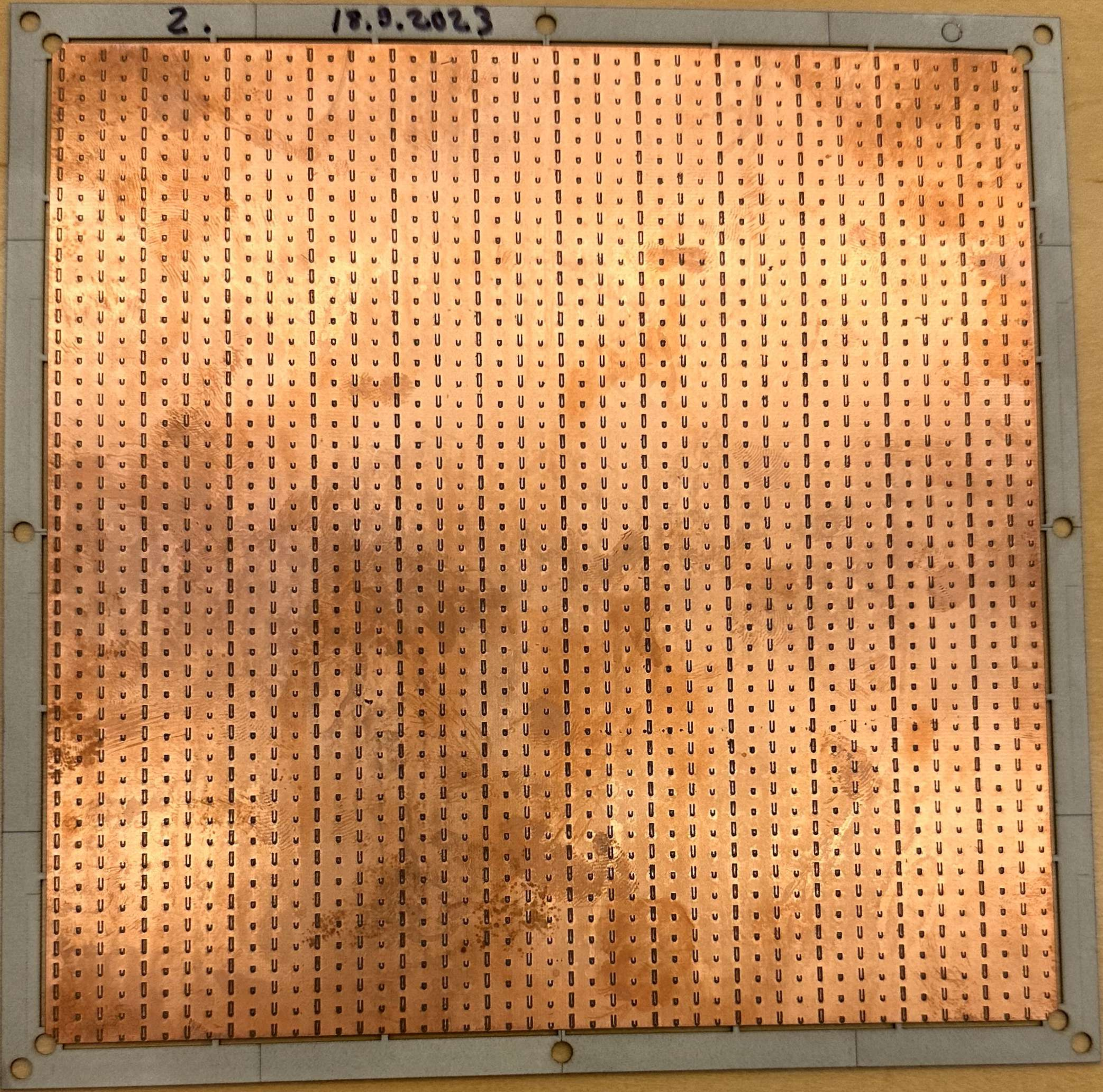}}
\caption{A manufactured $48\times48$-sized anomalous reflector (a) front, (b) back. \label{Fig:AR}}
\end{figure}

\begin{figure}[t]
\center 
\includegraphics[width=0.4\textwidth]{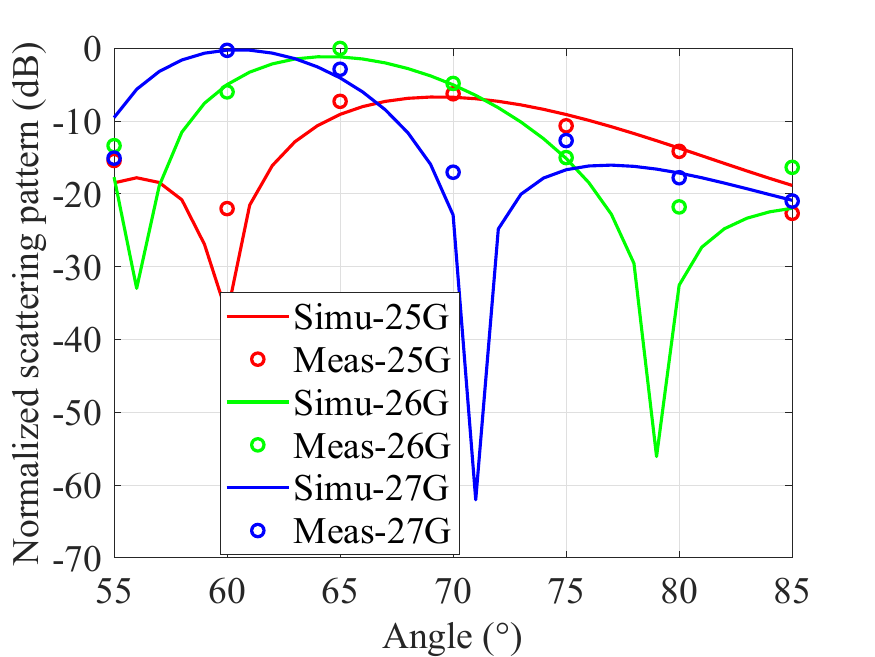}
\caption{Normalized scattering pattern of the $48\times48$-sized AR at different frequencies and angles from CST simulation and measurement.}
\label{Fig:pattern} 
\end{figure}

The manufactured 3-bit $48\times48$ square-shape array prototype is shown in Fig.~\ref{Fig:AR}, and its scattering property was verified through a measurement in an anechoic chamber in \cite{vuyyuru2024}. Now, we want to use this realistic reflector in our theoretical models and ray tracing tool to verify our implementations and communication models, as well as to evaluate the performance of this reflector in a real environment. Therefore, more measurement in a realistic scenario is needed. For convenience of our analysis, we reformulate the measured and simulated scattering pattern results for the $48\times48$-sized AR in Fig.~\ref{Fig:pattern}, as also seen in Fig.~7 in \cite{vuyyuru2024}. The measurement results are in good agreement with the CST simulation results in the main beam directions. It should be mentioned that the AR was originally only designed for $26$~GHz, but the results indicate it also works well at $25$ and $27$~GHz, with the main reflection direction shifted for $5\degree$. For example, the main reflection direction occurs at $70\degree, 65\degree,$ and $60\degree$ from $25$~GHz to $27$~GHz, respectively.

\section{Measurement System}
\label{sec:MeasSystem}
In this section, we perform over-the-air measurements in an auditorium with the manufactured $48\times48$-sized AR ($152.6$~mm$ \times 152.6$~mm) to evaluate its beamforming performance in the communication aspect, i.e., the large scale fading through the Tx-AR-Rx link. In addition, we combine four identical pieces of the $48\times48$-sized AR to form a large \ac{AR} with the size of $96\times96$ ($305.3$~mm$ \times 305.3$~mm). 
 
\begin{figure*}[!t]
\centering		
\subfigure[\label{Fig:3a}]{\includegraphics[width=0.715\textwidth]{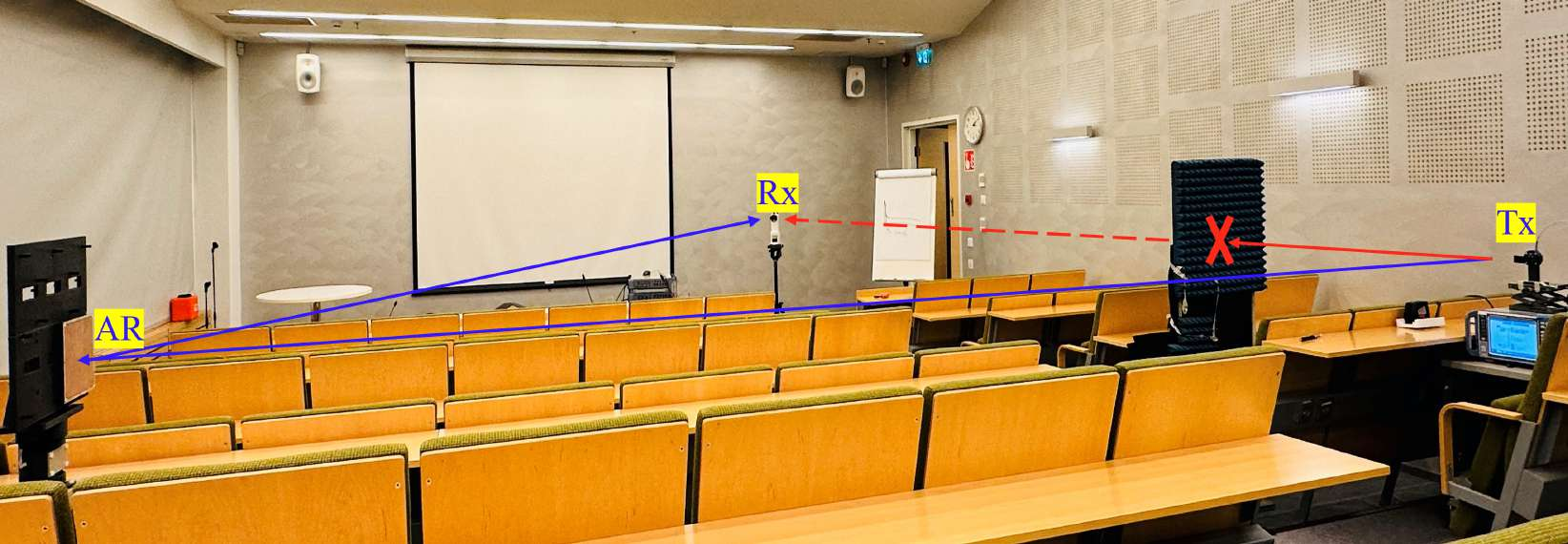}}
\subfigure[\label{Fig:3b}]{\includegraphics[width=0.139\textwidth]{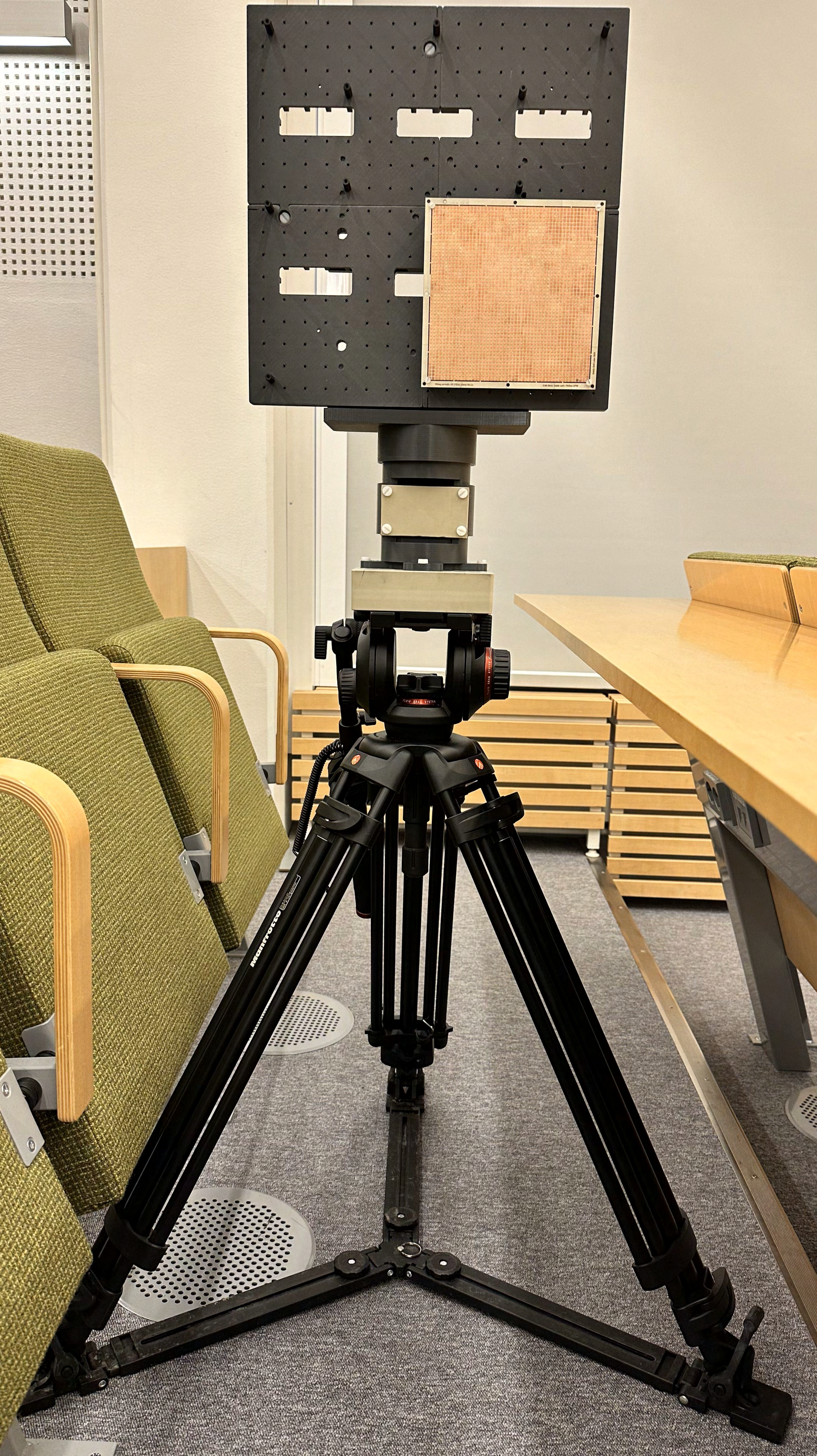}}
\subfigure[\label{Fig:3c}]{\includegraphics[width=0.133\textwidth]{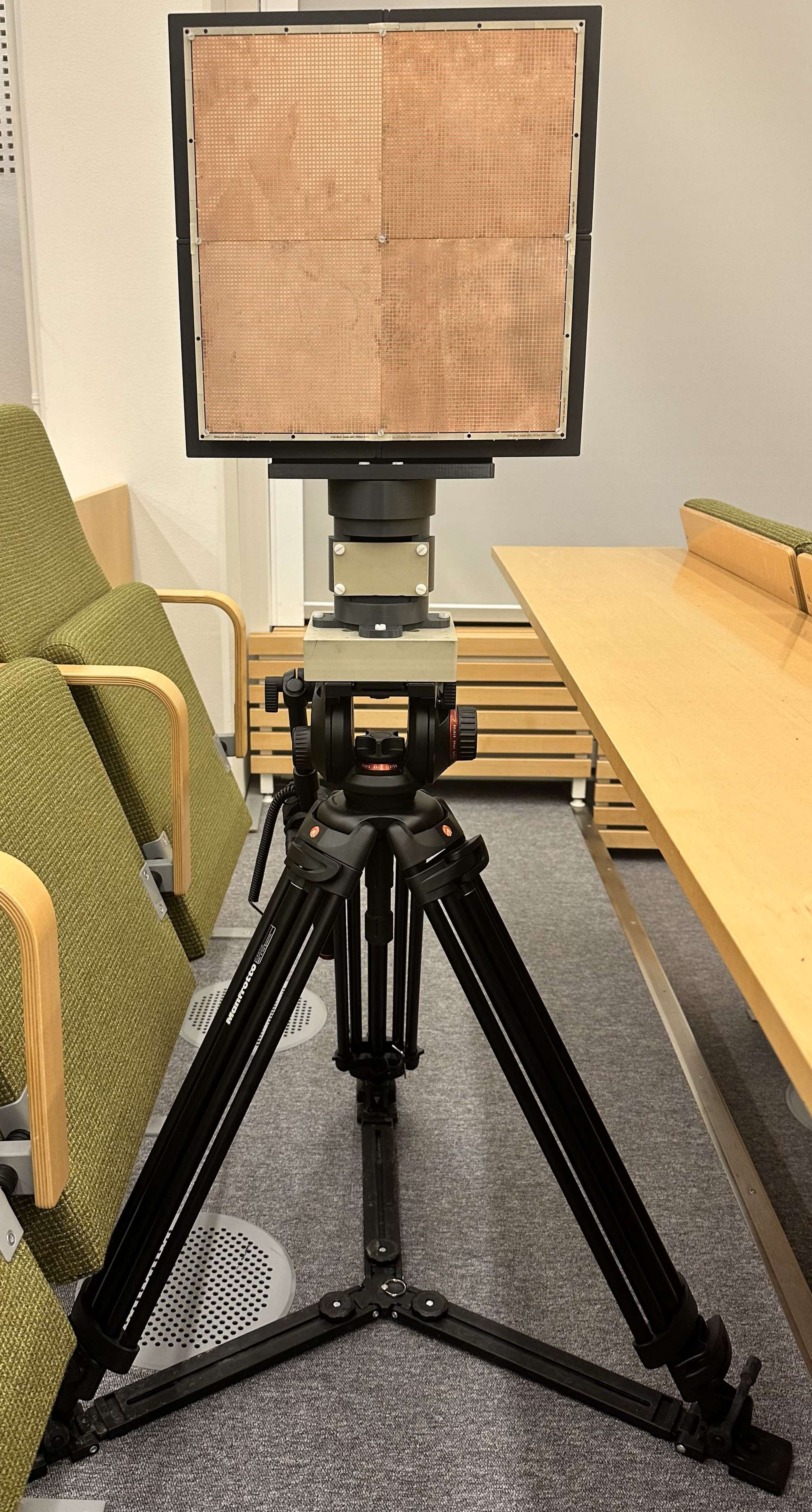}}
\caption{Measurement scenario and equipment.  (a) The measurement scenario in an auditorium (b) $48\times48$-sized AR ($152.6$~mm$\times 152.6$~mm), (c) $96\times96$-sized AR ($305.3$~mm$\times 305.3$~mm). \label{Fig:MeasScen}}
\end{figure*}

We use copper tape to glue the edges to ensure continuous alignment of elements with neighboring small ARs in the same plane, minimizing misalignment. The combined $96\times96$-sized AR is displayed in Fig.~\ref{Fig:3c}. We then repeat the measurement with the big AR in the auditorium to investigate its performance.

\subsection{Measurement scenario and setups}
\label{sec:MeasSet}
The measurement was conducted in a $14 \times 8 \times 3$ (m$^3$) sized auditorium at the Nokia Bell Labs office in Espoo, Finland. As shown in Fig.~\ref{Fig:MeasScen}, the Tx and Rx antennas are the same horn antennas. We fix the positions of the AR and the Tx antenna and move the Rx antenna to $55\degree, 60\degree, 65\degree, 70\degree, 75\degree, 80\degree,$ and $85\degree$ from the AR, respectively. The heights of the Tx, Rx antennas, and the AR center are $1.5$~m. The distance between the Tx and the AR is $5.5$~m, and the distance between the AR and the Rx antenna is always $7$~m. The estimated Fraunhofer far-field distance of the $48\times48$ and $96\times96$-sized ARs are $4.04$~m and $16.17$~m, respectively. Therefore, the $48\times48$-sized AR is in the far-field region of the Tx and the Rx antennas and the $96\times96$-sized AR is in the radiating near-field region. There are \ac{LoS} paths between the Tx and the AR and between the AR and the Rx antenna, while the direct link between the Tx and the Rx antennas is blocked by a wave absorber. Throughout the measurement campaign, we avoided moving objects in the room.

\begin{figure}[t]
\center 
\includegraphics[width=0.48\textwidth]{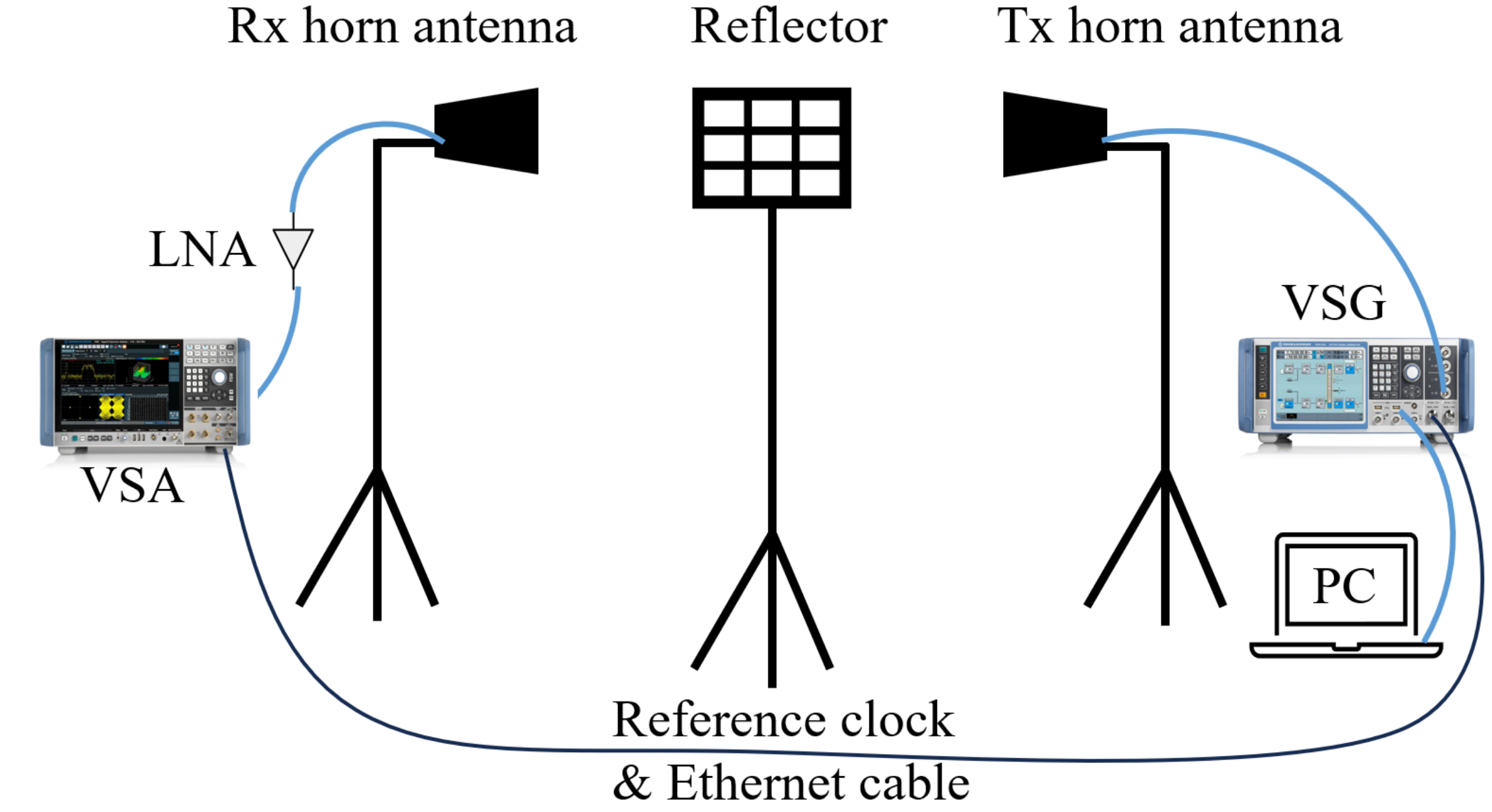}
\caption{Measurement setup.}
\label{Fig:MeasSet} 
\end{figure}

The measurement setup is shown in Fig.~\ref{Fig:MeasSet}. A Tx horn antenna is connected to an R\&S vector signal generator (VSG) SMW200A and then connected to a control computer. We generated a 5G NR PUSCH signal characterized by full resource blocks and $50\%$ duty cycle within a channel bandwidth of $400$~MHz or $100$~MHz and a subcarrier spacing of $120$~kHz for $26$~GHz mmWave band by the VSG and the Tx output power calibrated to $+6$~dBm. The Rx horn antenna is connected to a low-noise-amplifier (LNA) Miteq JS4 with a noise figure of $2.7$~dB; the LNA is then connected to an R\&S FSW vector signal analyzer (VSA) with a capture buffer length of $20.1$~ms. Vertically polarized Tx and Rx horn antennas were configured with a gain of $18$~dB with a beamwidth of $22\degree$. The VSA is connected to the VSG via an Ethernet cable and reference clock for synchronizing. The parameters of the devices are listed in Table~\ref{tab:Setup}. In our measurement, we transmit 16~QAM \ac{OFDM} modulated signals. 

Next, the \ac{EVM} and received power levels were measured through the Tx-AR-Rx link. \ac{EVM} is a communications systems metric to indicate the received signal quality throughout the measured channel bandwidth for evaluating the conditions for demodulation of the received signal. The received power defines the amount of power acquired by the user receiver in assessing the performance.
Since this kind of measurement takes a long time with the used instrumentation, we measure only at three distinct frequencies, $25, 26, $ and $27$~GHz with the modulated signals. For reference, we use continuous wave signal to sweep from $24.5$~GHz to $27.5$~GHz with a step of $0.25$~GHz to have proper frequency responses and a quicker measurement setup. 

\begin{table}[t]
\caption{MEASUREMENT PARAMETERS}
\label{tab:Setup}
\centering
\begin{tabular}{|c |c| c |}
\hline
  & Item & value \\ \hline
Transmit power  & $P_t$ & $6$~dBm \\ \hline
Cable loss at Tx side & $(-) L_t$  & $-2.5$~dB \\ \hline
Tx antenna gain & $G_t$ & $18$~dB \\ \hline
Rx antenna gain  & $G_r$  & $18$~dB  \\ \hline
LNA gain + cable loss at Rx side & $G_a$ & $19.9$~dB \\ \hline
Distance between the Tx and AR & $R_1$ & $5.5$~m \\ \hline
Distance between the Rx and AR & $R_2$ & $7$~m \\ \hline
\end{tabular}
\end{table}

\subsection{Path loss models}
\label{sec:Linkbudget}
Different path loss models for ARs have been discussed in~\cite{Hao2024,Hao2024Eucap}. For periodic metasurfaces, the distribution of reflected fields over all angles can be estimated based on the results of Section~II in \cite{Sergei2023}, also for imperfect ARs that produce some parasitic scattering into undesired Floquet modes. Propagation of signals due to reflections from some objects is conventionally modeled in terms of the bi-static scattering cross-section that can be determined based on the analytical model in  \cite{Sergei2023}. For reflections into the desired direction that model gives a very simple analytical formula, which we use as the first method in this work for a theoretical analysis of our manufactured AR. We denote the received power estimated by  Eq.~(38) in \cite{Sergei2023} as $P_1$. Adding the measurement setup parameters, we obtain the received power $P_r=P_1-L_t+G_a$ in dBm from method 1. 

While using the theory of \cite{Sergei2023} it is possible to estimate the bi-static cross section for any reflection direction, for integration of the RIS models into ray-tracing simulators it is more convenient to use the model for the scattering performance of AR in terms of the AR panel gains in the directions of illumination and observation~\cite{Hao2024Eucap, Hao2024}, which we call method 2. This approach is based on the Friis formula and gives the received power estimation
\begin{equation}\label{equ:mtd2}
 P_2 = \frac{ P_t G_t G_\text{rx} G_\text{tx} G_r \lambda^4}{(4\pi)^4(R_1 R_2)^2}.
\end{equation}
Here, $G_\text{rx}$ is the gain of the AR in the direction from AR to Tx, and $G_\text{tx}$ is the gain in the direction from AR to Rx. They are obtained from the CST simulation of the AR, where we first calculate the directivity of the AR (denoted as $D_i$) from the far-field pattern in CST simulation, then we use the directivity value to approximate the gain of the AR according to $G = e_{\text{cd}} D_i$~\cite{balanis2015antenna} and assume the panel efficiency $e_{\text{cd}} = 1$. 
This theoretical model combines analytical estimations of the propagation loss with the radiation pattern results from \ac{EM} simulations. By using the simulated AR gain at different angles, we can obtain received power results at the corresponding angles. Similarly, the received power from method 2 is calculated as $P_r=P_2-L_t+G_a$ in dBm. Note that Eq.~\eqref{equ:mtd2} is the same as Eq.~(6) in~\cite{Ellingson2021}, but here we use it for the whole RIS panel, and not for a single array element, as in~\cite{Ellingson2021}. Finally, we note that the product of the two gains of the AR panel is related to its bi-static scattering cross-section $\sigma$ as 
\begin{equation}
    G_\text{rx} G_\text{tx}=\frac{4\pi\sigma}{ \lambda^2},
\end{equation}
which is a function of both angles of incidence and observation. This equation establishes the relation between the results given by the two used methods.

\begin{figure}[t]
\center 
\includegraphics[width=0.48\textwidth]{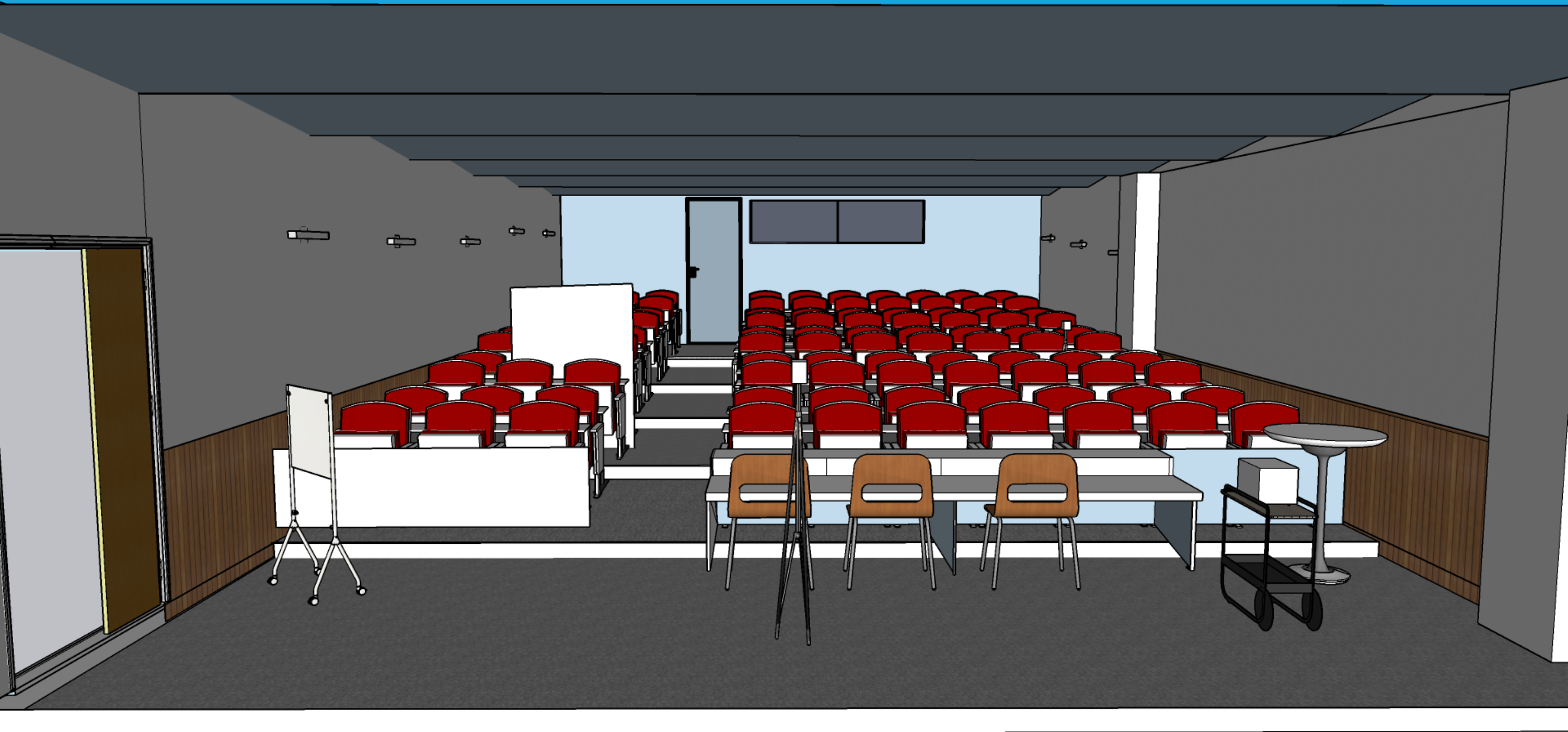}
\caption{A front view of the 3D auditorium model, which is replicated with a 1:1 ratio from the actual sizes.}
\label{Fig:room3D} 
\end{figure}

\subsection{Ray tracing model}
\label{sec:RTModel}
A ray tracing simulation tool accurately evaluates the wireless communication link performance in realistic scenarios. To reproduce the measurement scenario for ray tracing simulations, we first create a 3D model of the auditorium and objects inside with their actual dimensions in the software SketchUp, then import the 3D model to the MATLAB ray tracer for simulations. The front view of the 3D model is shown in Fig.~\ref{Fig:room3D}. We model a horn antenna with the maximum gain of $18$~dB as Tx and Rx antennas and import the radiation patterns of the AR from CST simulations to the ray tracer. 

Initially, CST simulations are conducted on a periodic (infinite) array of supercells. Subsequently, finite-size array scattering patterns are derived using the embedded element approximation in CST by applying an array factor calculation on the ``element pattern" of the simulated single supercell, as stated in \cite{vuyyuru2024}. The $+65\degree$ anomalous reflection scattering pattern with normal incidence is utilized when the AR functions as a Tx during the AR-Rx link. Conversely, the $0\degree$ scattering pattern is employed with $-65\degree$ incidence when the AR serves as an Rx in the Tx-AR link. The ray tracer cannot perform a direct link Tx-AR-Rx due to the definition of the AR as an antenna. Specifically, if the scattering pattern only reflects at a $+65\degree$ angle with normal incidence, the AR will not receive any signal from the Tx because the scattering pattern has a low value at a $0\degree$ angle. It is necessary to integrate two scattering patterns from CST into the ray tracer to address the problem. 

\begin{figure}[t]
\center 
\includegraphics[width=0.45\textwidth]{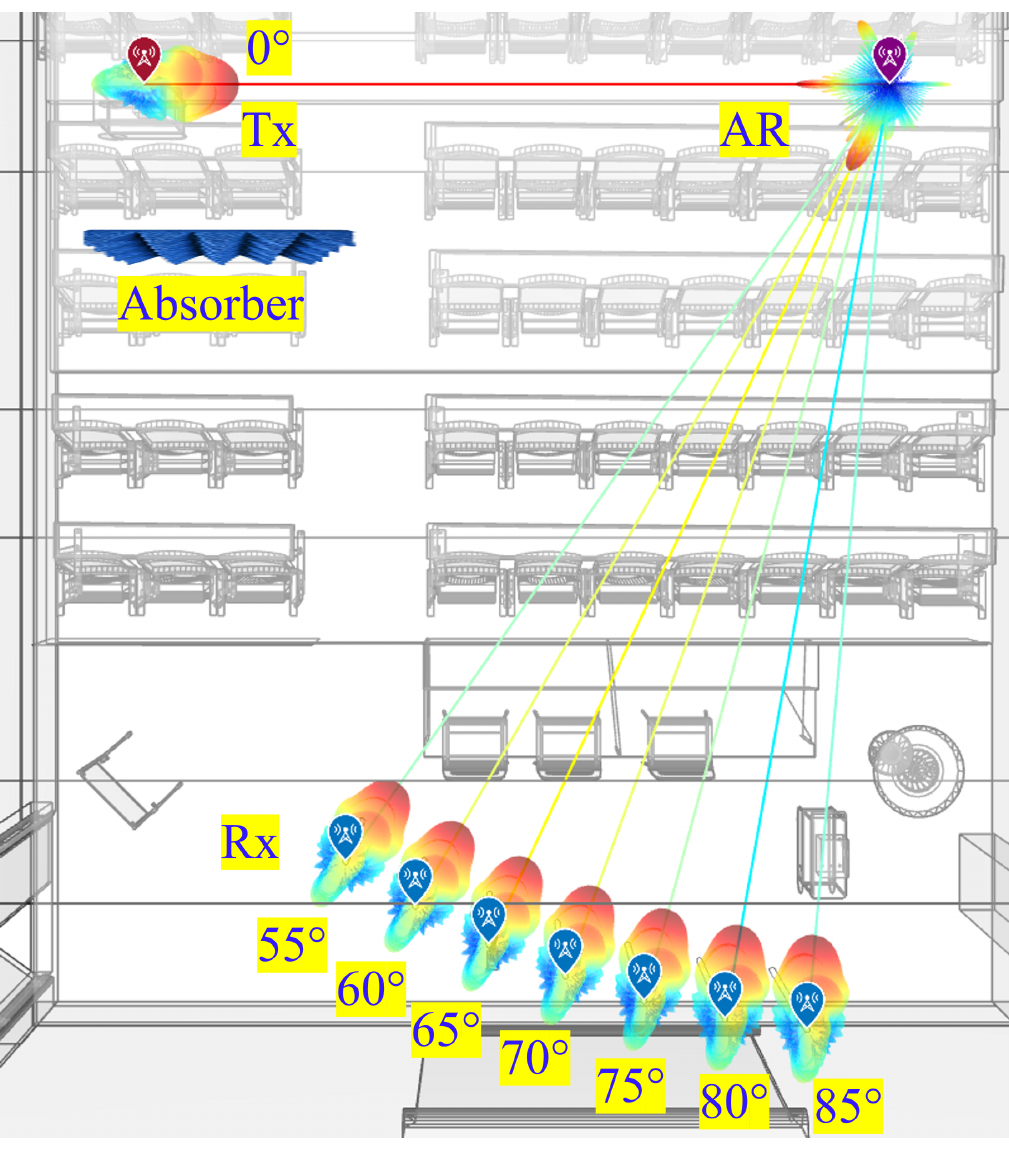}
\caption{The Tx, Rx, and AR positions and radiation patterns in the MATLAB ray tracer, as well as the LoS paths between them.}
\label{Fig:RTLoS} 
\end{figure}

We place the Rx antenna at different positions from $55\degree$ to $85\degree$ and orient the main beam of the Rx antenna pattern to always face the AR. In Fig.~\ref{Fig:RTLoS}, we plot the \ac{LoS} paths between the Tx and the AR, and between the AR and the Rx in the MATLAB ray tracer. The locations and heights of the Tx, Rx, and AR are the same as in the measurement. For ray tracing simulations with the two ARs, we set the material type of the auditorium as ``concrete" and set the reflection paths as zero, or three for comparison. When there is no reflection path, it will only simulate the LoS paths as shown in Fig.~\ref{Fig:RTLoS}, and the path loss calculation is based on the free space path loss model which is the same as from method 2. The zero-reflection case is used to compare the simulation results with the theoretical model, and to allow for a more practical propagation environment, we set three reflection paths to compare the results with measurement results.

\section{Measurement and Simulation Results}
\label{sec:Results}
In this section, we compare received power results through the AR-assisted communication links between the theoretical models, ray tracing simulations, and measurements. We analyze the AR's angular response across various Rx locations, evaluate frequency selectivity by varying frequencies, and assess spatial scalability performance through AR prototypes.

\subsection{Reference measurement of LoS links}
\label{sec:losmeas}

To obtain accurate measurement results and have a fair comparison with simulation results, we first conduct a reference measurement with only LoS links between the Tx and the Rx antenna. Next, we compare the measured results with theory to find out how much power difference we obtain between them. During the measurement, we fix the Tx antenna position and move the Rx antenna to $7$ different locations from $55\degree$ to $85\degree$ with a $5\degree$ step because the simulated half power beam width (HPBW) of the $48\times 48$-sized AR is $9\degree$, a $5\degree$ step is enough to catch the peak power. However, the HPBW of the $96\times 96$-sized AR is $5\degree$. Therefore, we add an additional angle of $62.5\degree$ for the $96\times 96$-sized AR measurement. At each Rx location, we orient the Tx and Rx antennas to let their main beam directions toward each other, and we measure the received power at the Rx antenna through the LoS path from the Tx antenna. We conduct the measurement first with 16~QAM modulated waves at $25, 26,$ and $27$~GHz, then we use continuous waves to sweep from $24.5$~GHz to $27.5$~GHz with $0.25$~GHz step. 

We denote the measured received power value as $P_m$. Then we calculate the free space path loss between the Tx and the Rx using the Friis formula $P_{\text{FSPL}} = \left(\frac{\lambda}{4\pi R_3}\right)^2$ in $W$, where $R_3$ denotes the distances between the Tx and the Rx antenna. The final calculated received power from theory is $P_{\text{theory}} = P_t+G_t+G_r+P_{\text{FSPL}}-L_t+G_a$ in dBm. The power difference between the theoretical and measurement results is $P_{\text{diff}} = P_{\text{theory}} - P_m$ in dB. For modulated-wave measurement, we only use the results at $25, 26,$ and $27$~GHz, while for continuous-wave measurement, we use values at all frequencies to calculate the power difference. 

\begin{table}[t]
\caption{POWER DIFFERENCES BETWEEN THEORY AND MEASUREMENTS}
\label{tab:Pdiff}
\centering
\begin{tabular}{|c|c|c|c|c|c|c|c|}
\hline
$P_{\text{diff}}$ (dB) & $55\degree$ & $60\degree$ & $65\degree$ & $70\degree$ &$75\degree$ & $80\degree$ & $85\degree$ \\ \hline
$25$~GHz & $0.87$  & $0.78$ & $0.77$ & $0.25$  & $0.14$  &$0.28$  & $1.52$\\ \hline
$26$~GHz & $1.46$  & $1.55$ & $1.08$ & $0.68$  & $0.26$ & $1.05$ & $0.28$\\ \hline
$27$~GHz & $1.04$  & $2.13$ & $1.56$  & $1.33$ & $1.79$ & $-0.04$ &$1.85$\\ \hline
\end{tabular}
\end{table}

In Table~\ref{tab:Pdiff}, we list an example of the $P_{\text{diff}}$ with modulated measurement results at the three frequencies. These small differences may come from the losses in the measurement system that we could not count in our model. The losses due to measurement devices may have a little deviation between the measurement and simulation. For instance, the antenna gain variability at different frequencies is about $\pm 0.5$~dB. The LNA gain and the Rx cable loss together have about $1.6$~dB variation, and the Tx cable loss has around $0.065$~dB variability in the frequency range. When we take the $P_{\text{diff}}$ into account and use it to correct the calculated and simulated results with the AR, we can eliminate the influence of these factors and obtain more accurate results between the simulation model and the measurement. 

\begin{figure}[t]
\centering		
\subfigure[\label{Fig:7a}]{\includegraphics[width=0.24\textwidth]{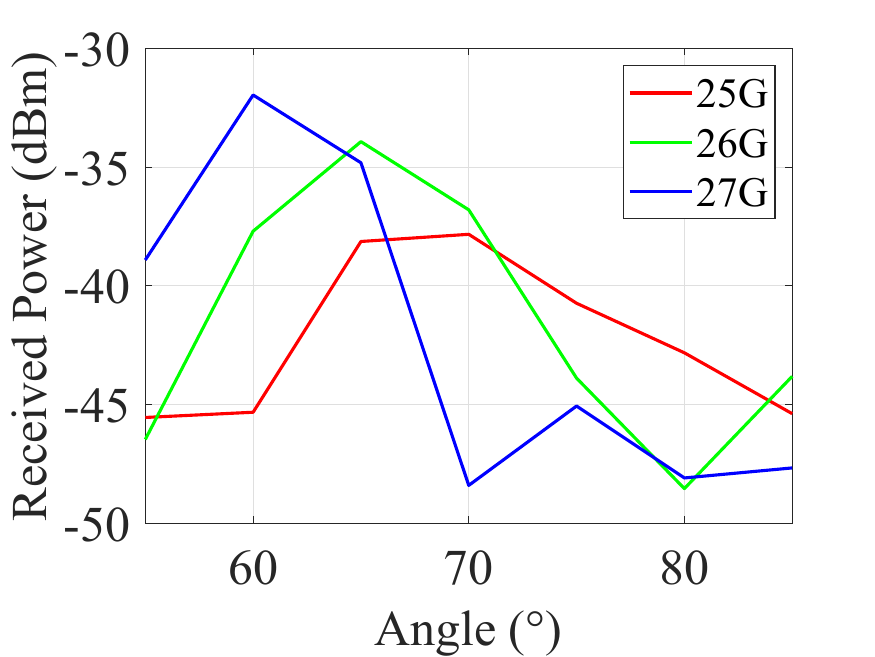}}
\subfigure[\label{Fig:7b}]{\includegraphics[width=0.24\textwidth]{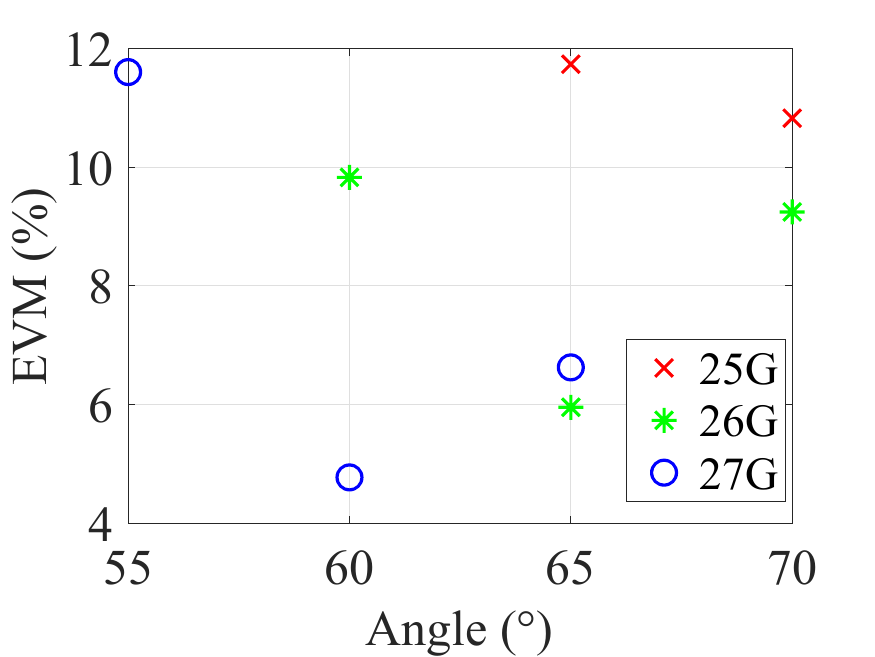}} 
\caption{Measured results of the $48\times 48$-sized AR with 16~QAM modulated waves. (a) received power vs. angle, (b) EVM vs. angle. \label{Fig:pwevmAR48}}
\end{figure}

\subsection{Results of the $48\times 48$-sized AR}
\label{sec:AR48}
For the $48\times 48$-sized AR, we measured the received power at $7$ Rx locations from $55\degree$ to $85\degree$ with a $5\degree$ step. Same as the previous LoS measurement, we perform measurements with 16~QAM modulated \ac{OFDM} waves with $400$~MHz bandwidth at the frequencies $25, 26,$ and $27$~GHz, and obtain the results of received power and \ac{EVM} that are displayed in Fig.~\ref{Fig:7a} and Fig.~\ref{Fig:7b}, respectively. Fig.~\ref{Fig:7b} only shows the EVM values that passed the frame-averaged EVM test with the 5G NR signal (EVM less than 12.5\% for 16QAM). We observe that from $25, 26,$ to $27$~GHz, the maximum received power increases while the minimum EVM decreases. The maximum received power and minimum EVM for $25, 26,$ and $27$~GHz occur at $70\degree, 65\degree$, and $60\degree$, respectively. The angles of the received power peaks at the three frequencies are consistent with the scattering pattern shown in Fig.~\ref{Fig:pattern}.

\begin{figure*}[t]
\centering		
\subfigure[\label{Fig:8a}]{\includegraphics[width=0.32\textwidth]{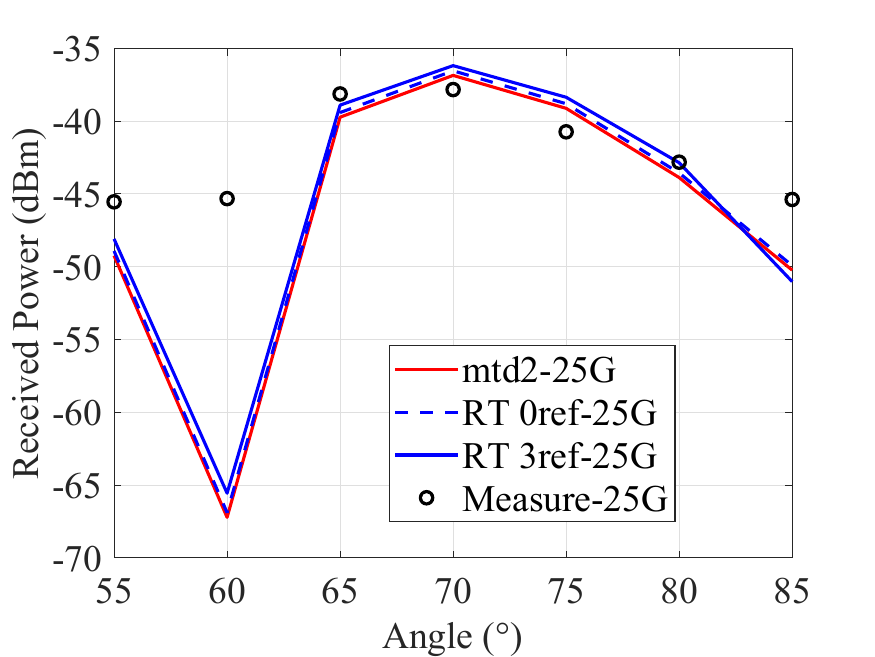}}
\subfigure[\label{Fig:8b}]{\includegraphics[width=0.32\textwidth]{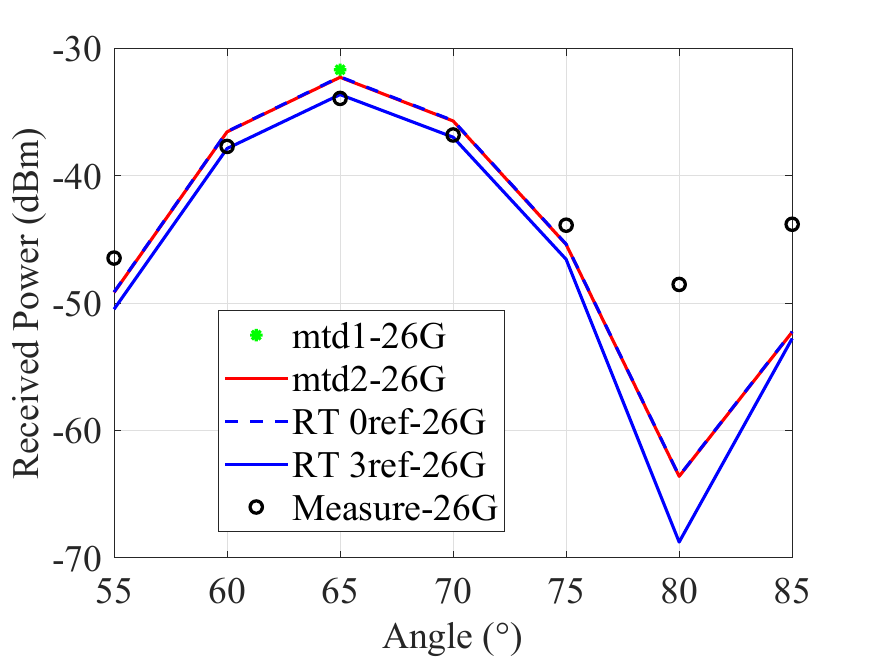}}
\subfigure[\label{Fig:8c}]{\includegraphics[width=0.32\textwidth]{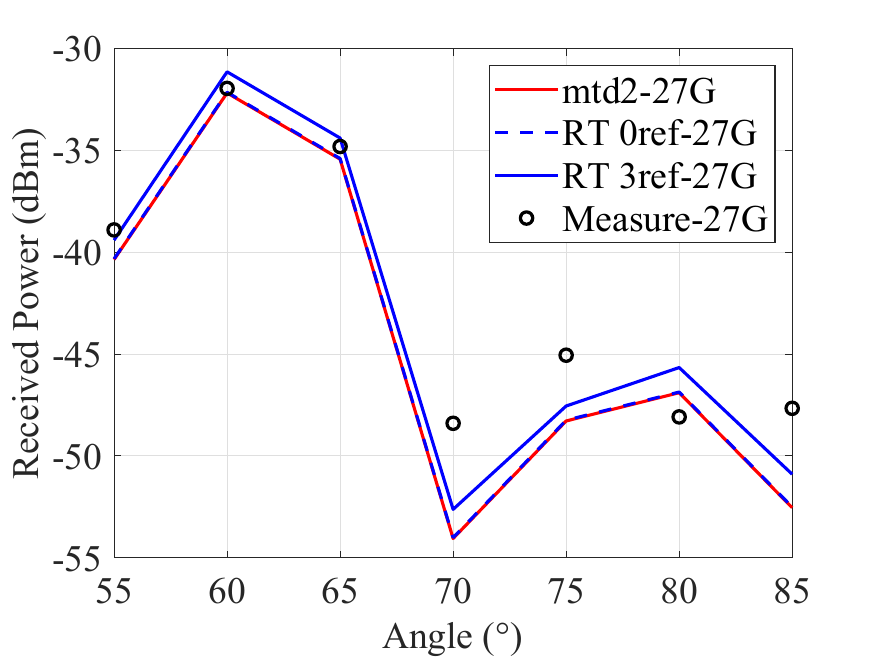}}
\caption{Results with the $48\times 48$-sized AR from corrected calculation and simulation, and from measurements using the modulated waves, (a) at $25$~GHz, (b) at $26$~GHz, (c) at $27$~GHz. \label{Fig:PW_correct}}
\end{figure*}	

\begin{figure*}[t]
\centering		
\subfigure[\label{Fig:9a}]{\includegraphics[width=0.32\textwidth]{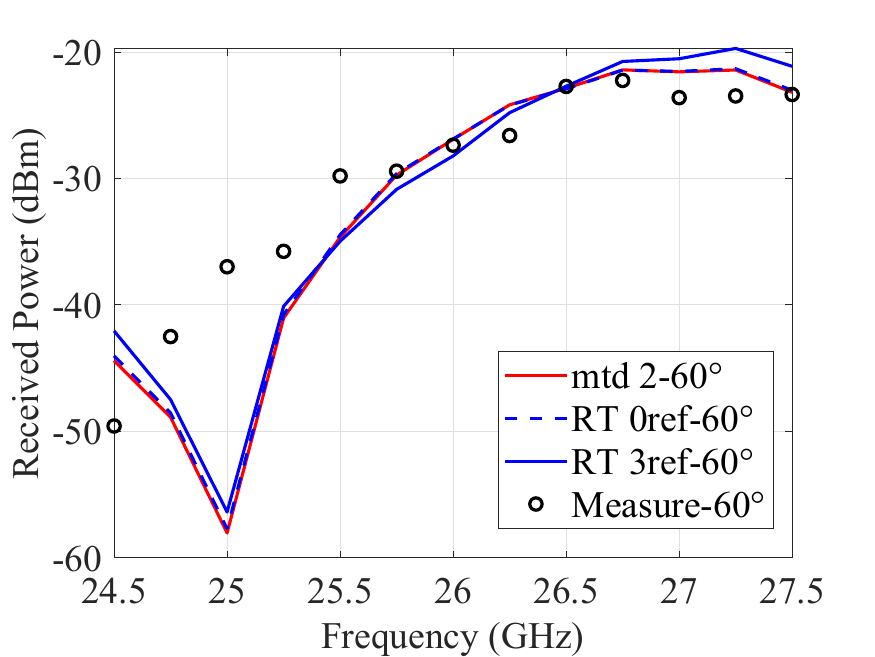}}
\subfigure[\label{Fig:9b}]{\includegraphics[width=0.32\textwidth]{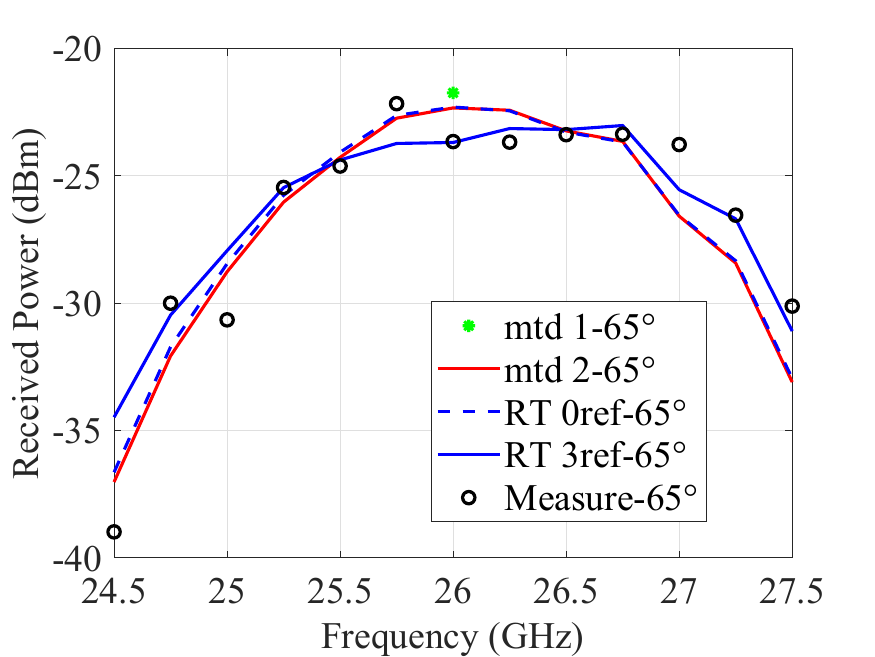}}
\subfigure[\label{Fig:9c}]{\includegraphics[width=0.32\textwidth]{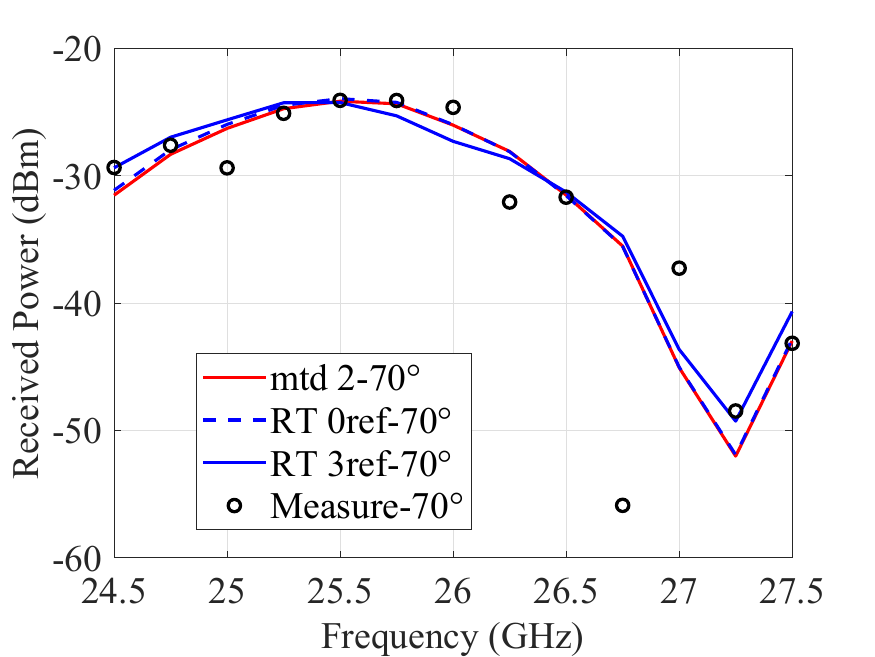}}
\caption{Results with the $48\times 48$-sized AR from corrected calculation and simulation, and from measurements using the continuous waves, (a) at $60\degree$, (b) at $65\degree$, (c) at $70\degree$. \label{Fig:PW_correct_cont}}
\end{figure*}	

The corrected calculation and simulation results for the AR-assisted links are obtained by $P_\text{correct} = P_r - P_{\text{diff}}$ with $P_r$ being the original power calculated from method 1 and method 2. The comparison of the corrected results and the measurement results are plotted in Fig.~\ref{Fig:PW_correct}. From these three figures, we observe that the results from method 2 and from ray tracing with zero reflection are consistent, which validates our AR implementation method in the ray tracer. Ray tracing with zero reflection and three reflections also does not show much difference in the received power. At $25$~GHz, the measurement and simulation results have very little difference at most angles, except at $60\degree$. This is highly related to the simulated and measured scattering pattern of the AR, as we see similar circumstances in Fig.~\ref{Fig:pattern}. At $26$~GHz, the measurement results at $60\degree, 65\degree,$ and $70\degree$ are almost the same as the ray tracing simulation results with three reflections. In addition, at $65\degree$ (the designed reflection angle), the theoretical results for perfect AR given by method 1 are very close to those given by method 2 and to the ray tracing results with zero reflection, which indicates that the 3-bit quantized lossy AR does not have much power loss at the designed reflection angle compared to a perfectly working lossless AR. At $27$~GHz, the simulation results and measurement results are almost the same at $55\degree, 60\degree$, and $65\degree$.  

To evaluate the results at more frequencies, we do the same correction procedure for the continuous-wave measurement results. Here we use the continuous-wave measured results $P_m$ and the theoretical results both from $24.5$~GHz to $27.5$~GHz and calculate the $P_{\text{diff}}$ between them, the corrected results $P_\text{correct}$ are from the updated $P_{\text{diff}}$. Since the nominal main beam direction of the AR is between $60\degree$ and $70\degree$ due to the frequency steering phenomenon, we observe the received power results at all $13$ frequencies and at $60\degree$, $65\degree$, and $70\degree$ which are shown in Figs.~\ref{Fig:9a}-\ref{Fig:9c}. Generally, we achieve good agreement between the simulations and measurement results. There are big differences only at a few frequency points, which is related to the AR scattering pattern difference between simulations and measurements. Since $65\degree$ is the designed reflection angle, we see that the agreement at this angle is better than for  $60\degree$ and $70\degree$. In addition, we find that the ``frequency steering'' phenomenon limits the use of this AR for signal bandwidths greater than approximately $1$~GHz. But for any individual center frequency within the given almost $3$~GHz range, we can operate. 

These observations confirm that our ray tracing model with the realistic AR agrees well with measurements in such an indoor scenario, even though the 3D model in the ray tracer does not fully replicate the auditorium since we can only set one material type for all objects in the MATLAB version up to 2024a, whereas the materials of different objects in the rooms are different in reality. Next, we can use this method to predict the communication performance of other ARs, especially those that are difficult to measure. For example, we could not measure the far-field scattering pattern of a $96\times96$-sized AR due to the limitation of the measurement setup, but we can use our models to predict the far-field performance of this AR.
\vspace{-1em}
\begin{figure}[h]
\centering		
\subfigure[\label{Fig:10a}]{\includegraphics[width=0.24\textwidth]{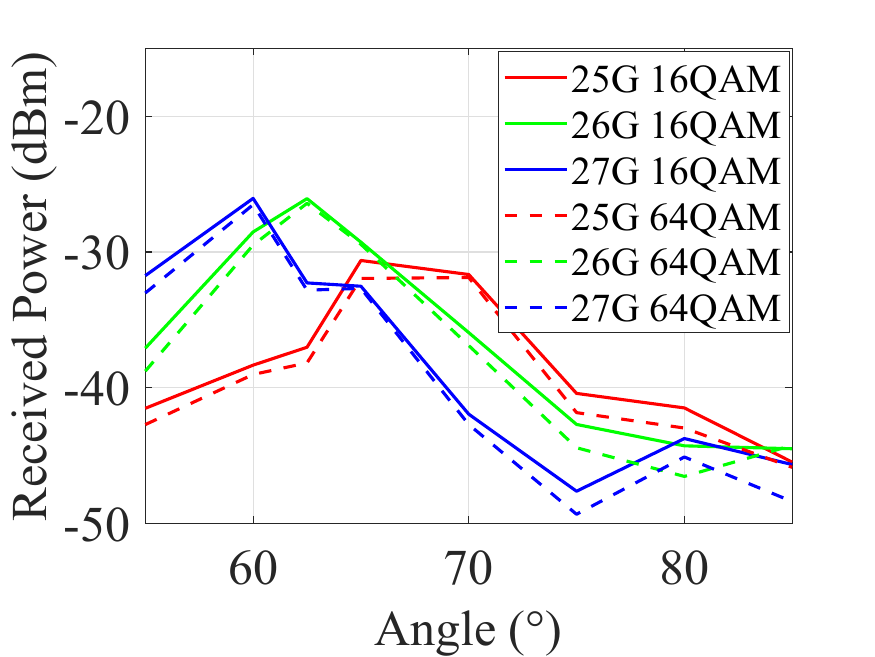}}
\subfigure[\label{Fig:10b}]{\includegraphics[width=0.24\textwidth]{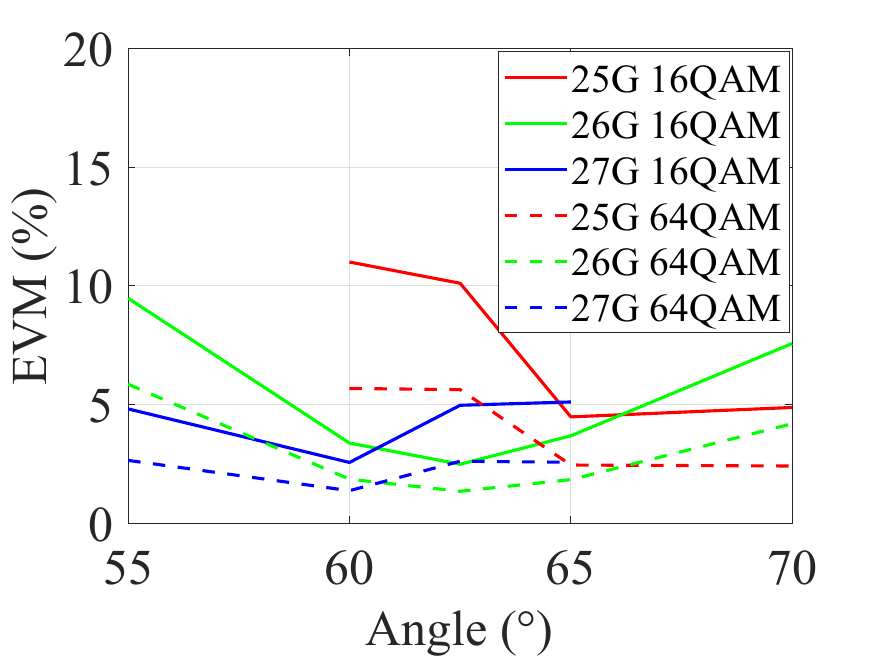}}
\caption{Measured results of the $96\times 96$-sized AR with 16~QAM modulated waves. (a) received power vs. angle, (b) EVM vs. angle.  \label{Fig:pwevmAR96}}
\end{figure}	

\subsection{Results for the $96\times 96$-sized AR}
\label{sec:AR96}
In this section, we analyze the results for the $96\times 96$-sized AR which is a tiled combination of four pieces of the $48\times 48$-sized AR. Since the size of the AR becomes four times larger than that of the small AR, the far-field distance of the big AR is not fulfilled anymore in our measurement. From the general properties of reflecting panels and  \cite{vuyyuru2024} we know that the main scattering beam of a larger AR is narrower than a smaller AR, and the scattering pattern of a larger AR is more sensitive in terms of the angles. Therefore, we repeat all the previous measurements for the big AR at the previous $7$ angles and additionally at $62.5\degree$, i.e., at $8$ angles in total for this AR measurement and simulations. In addition to 16~QAM modulated waves with $400$~MHz bandwidth, we also did measurements using 64~QAM modulated waves with $100$~MHz bandwidth. The measurement procedures are the same as those for the $48\times 48$-sized AR, i.e., with modulated waves and continuous waves, correcting the calculation and simulation results with the power difference $P_{\text{diff}}$. 

\begin{figure*}[t]
\centering		
\subfigure[\label{Fig:11a}]{\includegraphics[width=0.32\textwidth]{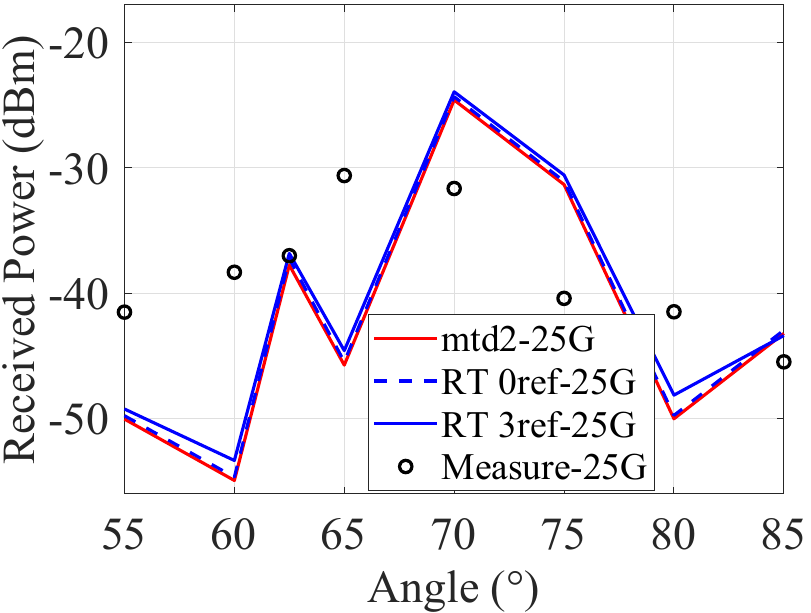}}
\subfigure[\label{Fig:11b}]{\includegraphics[width=0.32\textwidth]{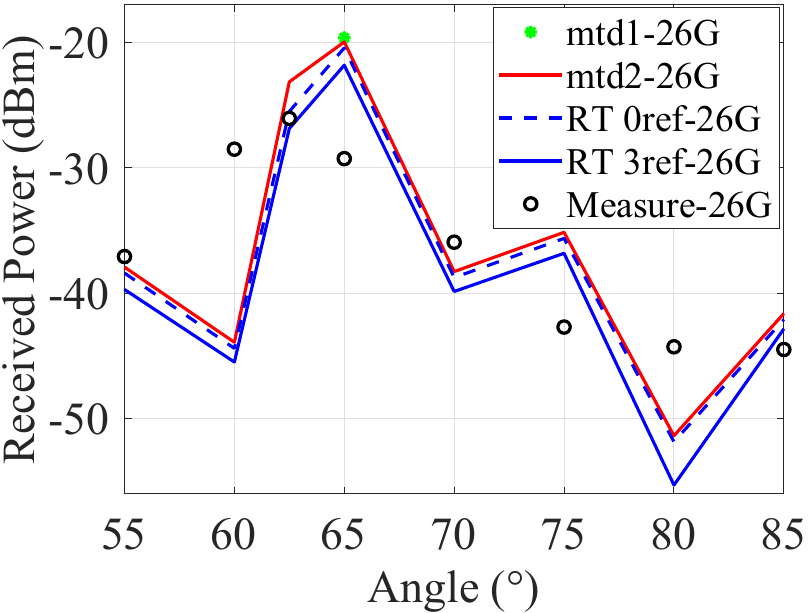}}
\subfigure[\label{Fig:11c}]{\includegraphics[width=0.32\textwidth]{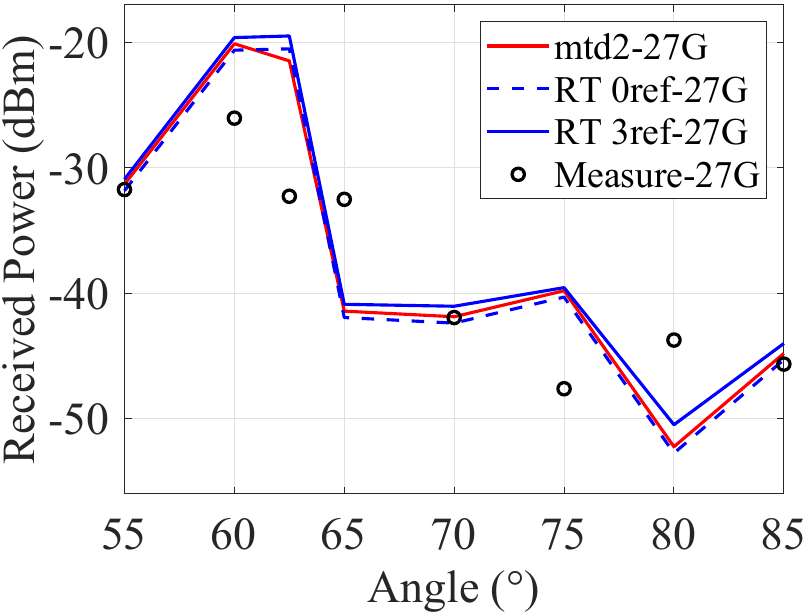}}
\caption{Results with the $96\times 96$-sized AR from corrected calculation and simulation, and from measurements using the modulated waves, (a) at $25$~GHz, (b) at $26$~GHz, (c) at $27$~GHz. \label{Fig:PW_correct96}}
\end{figure*}	

\begin{figure}[t]
\centering		
\subfigure[\label{Fig:12a}]{\includegraphics[width=0.24\textwidth]{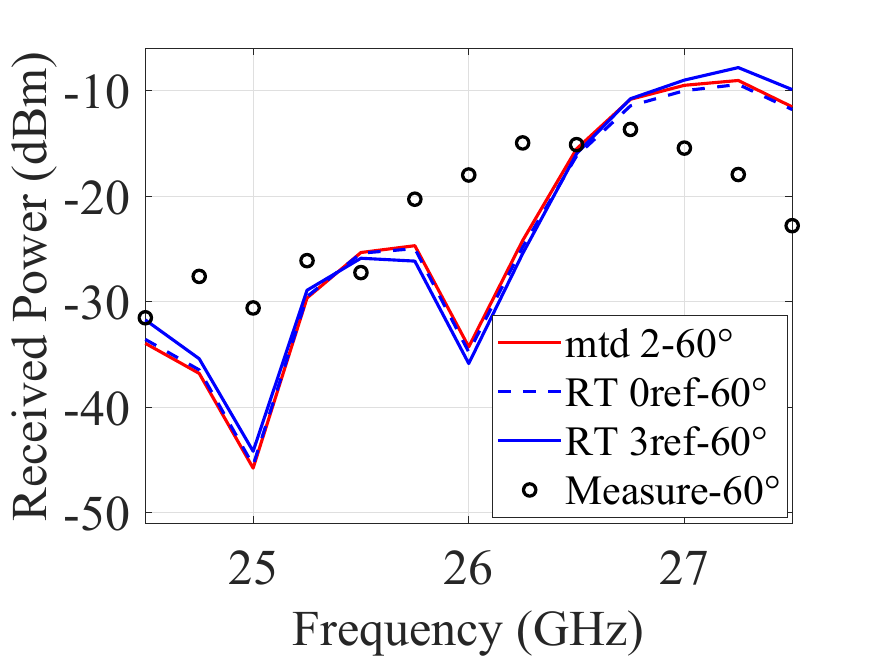}}
\subfigure[\label{Fig:12b}]{\includegraphics[width=0.24\textwidth]{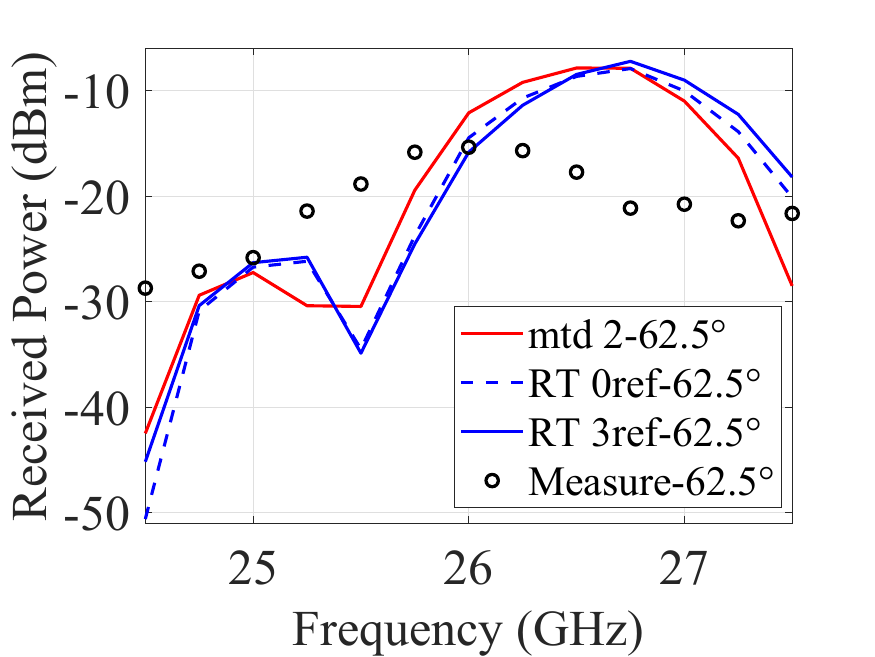}}
\subfigure[\label{Fig:12c}]{\includegraphics[width=0.24\textwidth]{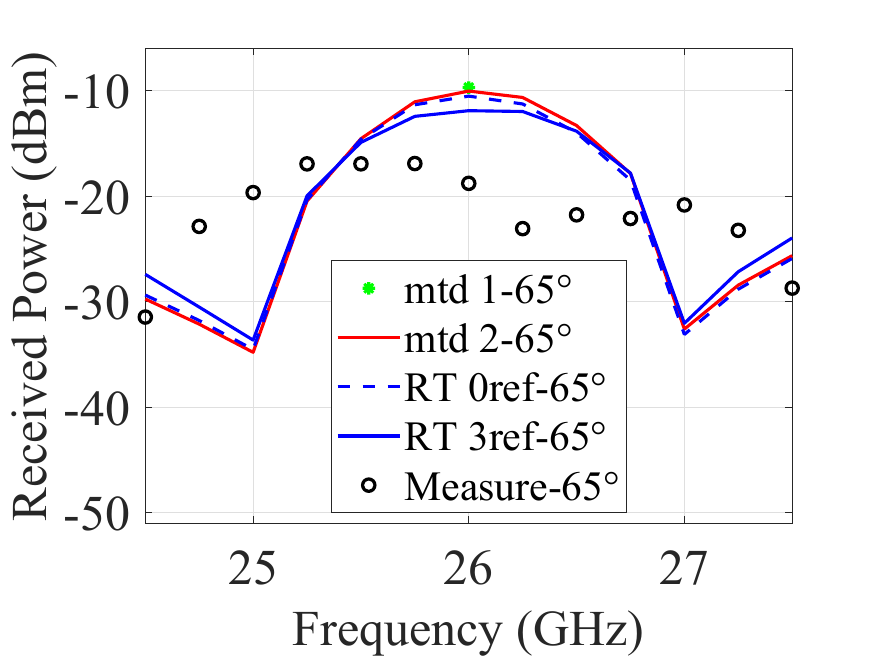}}
\subfigure[\label{Fig:12d}]{\includegraphics[width=0.24\textwidth]{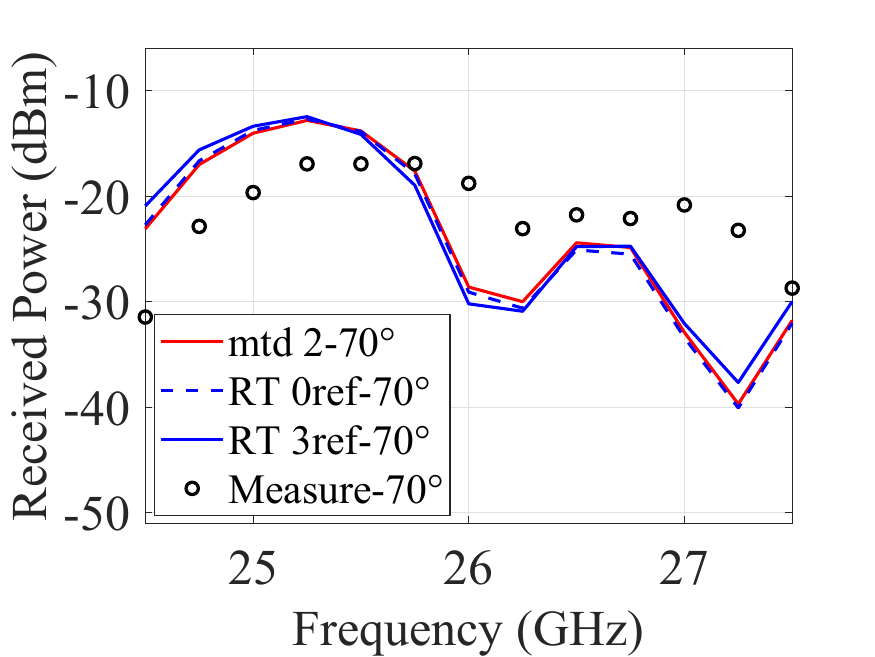}}
\caption{Results with the $96\times 96$-sized AR from corrected calculation and simulation, and from measurements using continuous waves, (a) at $60\degree$, (b) at $62.5\degree$, (c) at $65\degree$, (d) at $70\degree$. \label{Fig:PW_correct96_cont}}
\end{figure}	

The received power and EVM results from the big AR are shown in Fig.~\ref{Fig:10a} and Fig.~\ref{Fig:10b}, respectively. We only plot the EVM results that passed the EVM tests. These results indicate that the maximum power for $25, 26,$ and $27$~GHz appear at $65\degree$, $62.5\degree$, and $60\degree$, respectively. The EVM results are consistent, i.e., the minimum EVM results for the big AR also appear at the same angles for the three frequencies. The main beam direction moves to a smaller angle when increasing the frequency, which is the same as for the small AR. The only difference is that the peak power from the small AR appears at $70\degree$, $65\degree$, and $60\degree$ for $25, 26,$ and $27$~GHz, respectively. It also shows that the big AR works well with both 16~QAM and 64~QAM modulation schemes. When changing from 16~QAM to 64~QAM, it can achieve higher data rates and throughput. From the comparison of the results between the two modulation schemes, the 64~QAM modulation results in lower EVM and received power than using 16~QAM, but the differences are not very significant. Therefore, we only compare the simulation and calculation results with 16~QAM measurements in the following process.

The received power results from the corrected calculation and simulations, as well as from measurements using modulated waves are shown in Fig.~\ref{Fig:PW_correct96}. Fig.~\ref{Fig:11a}-\ref{Fig:11c} display the results at $25, 26,$ and $27$~GHz, respectively. These three figures show different beam shapes compared to the $48\times48$-sized AR. The measured power peak appears on $62.5\degree$, while the simulated peak is still on $65\degree$ at $26$~GHz. However, an interesting finding is that at some angles, the simulation and measurement results match very well, such as the power in $62.5\degree$ at $25$~GHz and $26$~GHz. Since the beam width of the large AR is much narrower than that of the small AR, the simulation and measurement results have generally larger differences in different angles. These big differences may be due to the following reasons: 1) the accuracy of measurement is reduced due to the more sensitive angular response of the large AR, and the angle samples for measurement are too sparse for this AR; 2) the simulated AR gain used in \eqref{equ:mtd2} and the radiation pattern used in ray tracing simulations are from a $96\times96$ complete panel without discontinuity, while the measured AR is a manually combined one. Any misalignment of the unit cells from the four small ARs will reduce the total scattering efficiency of the big AR, and the copper glue at the back of the ARs may also influence reflections and the frequency selectivity of the AR; 3) the analytical expression \eqref{equ:mtd2} and the ray tracing simulations give far-field results for the big AR while the measured AR is in the near-field, thus the comparison between them does not match very well. 

Similar to the $48\times 48$-sized AR, we compare the results corrected by continuous wave measurement. The results at the $13$ frequencies are presented in Fig.~\ref{Fig:PW_correct96_cont}, which again show significant differences between the simulation and measurements. These results reveal that the big AR still works at $26$~GHz, but the main beam has a slight shift, and the working bandwidth for this AR is approximately $1$~GHz. It would be interesting to also compare the performance of this combined AR with a manufactured $96\times96$-sized complete AR and measure the scattering pattern of the big AR when the setup conditions allow. This will help us find out whether the differences come from the manual combining of the four ARs or from the radiation pattern differences between simulations and measurements. This work is postponed to the future as a next step.

\section{Conclusion}
\label{sec:Conclusion}
In this work, we introduce and study a manufactured AR that is designed to reflect normally incident plane waves toward $65\degree$ at $26$~GHz. The scattering pattern of this AR from CST simulation and measurement is consistent in the main reflection directions. We then conduct over-the-air measurements in an auditorium with the $48\times 48$-sized AR, to measure the received power at an Rx antenna through the Tx-AR-Rx link. In addition, we calculate the received power from the AR-assisted link theoretically. These results are compared with the ray tracing simulations, where we implement an AR model into the MATLAB ray tracer and simulate the actual complex propagation scenario as in the measurement. The calculated and simulated results show good agreement with the measurement results over different frequencies and different angles. 

In addition, we combine four pieces of the $48\times 48$-sized AR to form a $96\times 96$-sized AR and measure the received power at the Rx through anomalous reflection in this combined AR. Results show that the peak received power at observed angles shifts to $62.5\degree$ at $26$~GHz with the combined AR from measurement. The frequency response of this AR is not as good as that of the small AR, but the simulation and measurement results show good agreement at several angles and frequencies. It is proved that our method of implementing an AR in the MATLAB ray tracer is correct and that it can be extended to any kind of AR as long as the scattering pattern of the AR is available. Our designed AR has a wideband property and can work well within a $2.75$~GHz band. The designed AR also has good spatial scalability. When combining multiple smaller pieces of ARs to form a bigger AR, we observe that the big AR can achieve good performance with little deviation from the design goals. This property offers a convenient alternative to completely redesigning a new AR with a bigger size.

\ifCLASSOPTIONcaptionsoff
  \newpage
\fi
\bibliographystyle{IEEEtran}
\bibliography{IEEEabrv,Bibliography}

\end{document}

%% file: Acronyms.tex
\begin{acronym}[DSTTDSGRC]
\setlength{\itemsep}{-3pt}
\acro{BS}{base station}
\acro{ECDF}{empirical cumulative distribution function}
\acro{EM}{electromagnetic}
\acro{LoS}{line-of-sight}
\acro{SISO}{single-input single-output}
\acro{SLS}{system-level simulator}
\acro{AR}{anomalous reflector}
\acro{OFDM}{orthogonal frequency-division multiplexing}
\acro{EVM}{error vector magnitude}
\end{acronym}